\documentclass[%
 reprint,
superscriptaddress,
 amsmath,amssymb,
 aps,
 prl,
]{revtex4-2}
\usepackage{graphicx}  % needed for figures
\usepackage{mathrsfs}
\usepackage{multirow}
\usepackage{siunitx}
\usepackage{lineno}
\usepackage{soul}
\usepackage[normalem]{ulem}
\usepackage{booktabs}
\usepackage[colorlinks=true,allcolors=blue]{hyperref}

\usepackage{enumitem}
\newlist{todolist}{itemize}{2}
\setlist[todolist]{label=$\square$}
\usepackage{pifont}

\begin{document}

\title{Search for MeV-Scale Axionlike Particles and Dark Photons with PandaX-4T}

% !TEX root = ../main.
%updated on 2024/03/25

\def\shKeyLab{School of Physics and Astronomy, Shanghai Jiao Tong University, Key Laboratory for Particle Astrophysics and Cosmology (MoE), Shanghai Key Laboratory for Particle Physics and Cosmology, Shanghai 200240, China}
\def\scKeyLab{Jinping Deep Underground Frontier Science and Dark Matter Key Laboratory of Sichuan Province}
\def\BUAA{School of Physics, Beihang University, Beijing 102206, China}
\def\BUAACenter{Peng Huanwu Collaborative Center for Research and Education, Beihang University, Beijing 100191, China}
\def\BUAALab{Beijing Key Laboratory of Advanced Nuclear Materials and Physics, Beihang University, Beijing, 102206, China}
\def\SCNT{Southern Center for Nuclear-Science Theory (SCNT), Institute of Modern Physics, Chinese Academy of Sciences, Huizhou 516000, China}
\def\USTClab{State Key Laboratory of Particle Detection and Electronics, University of Science and Technology of China, Hefei 230026, China}
\def\USTCdep{Department of Modern Physics, University of Science and Technology of China, Hefei 230026, China}
\def\BUAALab{International Research Center for Nuclei and Particles in the Cosmos \& Beijing Key Laboratory of Advanced Nuclear Materials and Physics, Beihang University, Beijing 100191, China}
\def\pku{School of Physics, Peking University, Beijing 100871, China}
\def\YaLongSD{Yalong River Hydropower Development Company, Ltd., 288 Shuanglin Road, Chengdu 610051, China}
\def\IAP{Shanghai Institute of Applied Physics, Chinese Academy of Sciences, 201800 Shanghai, China}
\def\CHEPpku{Center for High Energy Physics, Peking University, Beijing 100871, China}
\def\SDUdep{Research Center for Particle Science and Technology, Institute of Frontier and Interdisciplinary Science, Shandong University, Qingdao 266237, Shandong, China}
\def\SDUlab{Key Laboratory of Particle Physics and Particle Irradiation of Ministry of Education, Shandong University, Qingdao 266237, Shandong, China}
\def\UMD{Department of Physics, University of Maryland, College Park, Maryland 20742, USA}
\def\TDLee{New Cornerstone Science Laboratory, Tsung-Dao Lee Institute, Shanghai Jiao Tong University, Shanghai 201210, China}
\def\MESJTU{School of Mechanical Engineering, Shanghai Jiao Tong University, Shanghai 200240, China}
\def\SYU{School of Physics, Sun Yat-Sen University, Guangzhou 510275, China}
\def\SYUSFI{Sino-French Institute of Nuclear Engineering and Technology, Sun Yat-Sen University, Zhuhai, 519082, China}
\def\NKU{School of Physics, Nankai University, Tianjin 300071, China}
\def\YTU{Department of Physics, Yantai University, Yantai 264005, China}
\def\FDU{Key Laboratory of Nuclear Physics and Ion-beam Application (MOE), Institute of Modern Physics, Fudan University, Shanghai 200433, China}
\def\USST{School of Medical Instrument and Food Engineering, University of Shanghai for Science and Technology, Shanghai 200093, China}
\def\SJTUSC{Shanghai Jiao Tong University Sichuan Research Institute, Chengdu 610213, China}
\def\SPEIT{SJTU Paris Elite Institute of Technology, Shanghai Jiao Tong University, Shanghai, 200240, China}
\def\NNU{School of Physics and Technology, Nanjing Normal University, Nanjing 210023, China}
\def\SYSUzhuhai{School of Physics and Astronomy, Sun Yat-Sen University, Zhuhai 519082, China}
\def\CDUT{College of Nuclear Technology and Automation Engineering, Chengdu University of Technology, Chengdu 610059, China}

\affiliation{\TDLee}
%\author{Abdusalam Abdukerim}\affiliation{\shKeyLab}
\author{Tao Li}\affiliation{\SYUSFI}\affiliation{\SPEIT}
\author{Zihao Bo}\affiliation{\shKeyLab}
\author{Wei Chen}\affiliation{\shKeyLab}
\author{Xun Chen}\affiliation{\TDLee}\affiliation{\shKeyLab}\affiliation{\SJTUSC}\affiliation{\scKeyLab}
\author{Yunhua Chen}\affiliation{\YaLongSD}\affiliation{\scKeyLab}
%\author{Chen Cheng}\affiliation{\SYU}
\author{Zhaokan Cheng}\affiliation{\SYUSFI}
\author{Xiangyi Cui}\affiliation{\TDLee}
\author{Yingjie Fan}\affiliation{\YTU}
\author{Deqing Fang}\affiliation{\FDU}
\author{Zhixing Gao}\affiliation{\shKeyLab}
% ...
\collaboration{PandaX Collaboration}
\noaffiliation

\date{\today}
\begin{abstract}
Axionlike particles (ALPs) and dark photons (DPs) are viable dark matter particle candidates.
We have searched for possible ALP/DP signals in the PandaX-4T liquid xenon detector using 440 kg$\cdot$yr of data.
A binned likelihood fit is constructed to search for possible mono-energetic peaks induced by the absorption processes between ALPs/DPs and atomic electrons of xenon.
A detailed temporal model of decays associated with xenon isotopes is introduced to constrain the number of background events.
No signal excess over background expectations is observed, and we have established the most stringent exclusion limits for most ALP/DP masses across the range of 150 keV/$c^2$ to 1 MeV/$c^2$. 
The improvement is particularly significant within the mass range of \text{150-400}~keV/$c^2$,
with the average factor of 3.5 compared to previous results.
\end{abstract}

\maketitle
% \linenumbers

Astronomical and cosmological observations have provided compelling evidence for the existence of dark matter~(DM)~\cite{Planck:2018vyg, Bergstrom:2012fi, Feng:2010gw}, which is crucial for understanding the evolution of the universe.
For the last few decades, many terrestrial experiments worldwide have been dedicated to the search for DM, with a particular emphasis on weakly interacting massive particles (WIMPs)~\cite{Lee:1977ua, Goodman:1984dc, Bertone:2004pz, Liu:2017drf}, a prevailing candidate of cold dark matter~(CDM).
However, no conclusive signals from WIMPs have been detected so far.
On the other hand, there are some anomalies observed in the small-scale structure in galaxies which seem inconsistent with simulations within the CDM framework~\cite{Moore:1999gc,Klypin:1999uc,Moore:1999nt,Boylan-Kolchin:2011qkt,Weinberg:2013aya}.
This has prompted increased interest in alternative models involving lighter DM particles with weaker couplings to standard model~(SM) particles~\cite{Hall:2009bx, Essig:2011nj, Knapen:2017xzo}.

Among these models, the axionlike particles (ALPs) and dark photons (DPs), also referred to as bosonic super-WIMPs~\cite{Marsh:2015xka, An:2014twa, Pospelov:2008jk}, are of experimental interest. 
In contrast with the elastic scattering of WIMP with nucleus or electrons, they can be searched via unique absorption signals~\cite{Pospelov:2008jk}.
For example, the axioelectric effect, analogous to the photoelectric effect, will lead to the absorption of ALPs by the detector target with the energy transferred to one of the atomic electrons, producing a mono-energetic signal at the rest mass of ALPs.
The absorption cross section $\sigma_\mathrm{ALP}$ equals to $(3m_a^2c/16 \pi \alpha v m_e^2)\cdot g_{ae}^2\cdot \sigma_{pe}$~\cite{Pospelov:2008jk},
in which $m_a$ ($m_e$) is the mass of the ALP (electron), $\alpha$ is the fine-structure constant, $v$ is the velocity of the incoming ALP, $g_{ae}$ is the dimensionless coupling constant between the electrons and ALP, and $\sigma_{pe}$ is the photoelectric effect cross section for a photon with an energy of $m_a$.
Similarly, the cross section for DP is
$\sigma_\mathrm{DP} = (e^2c/4\pi\alpha v) \cdot \kappa^2 \cdot \sigma_{pe}$,
where $\kappa$ is the kinetic mixing constant between the DP and the real photon. 
Assuming that ALPs or DPs consist of all the DM in our galaxy with a density of 0.3 GeV/cm$^3$, the corresponding event rate in a terrestrial detector can be obtained as
\begin{equation}
\begin{aligned}
 R_\mathrm{ALP} & = \frac{1.47 \times 10^{19}}{A} g_{ae}^2 \cdot m_a \sigma_{pe}\,~[\mathrm{kg^{-1} d^{-1}}] \\
 R_\mathrm{DP} & = \frac{4.7 \times 10^{23}}{A} \frac{(e\kappa)^2}{4\pi\alpha} \frac{\sigma_{pe}}{m_d} \, ~[\mathrm{kg^{-1} d^{-1}}],
 \label{eq:event_rate}
\end{aligned}
\end{equation}
respectively, where $A$ represents the atomic mass of the absorbing atoms in the detector. 
The ALP (DP) mass $m_a$ ($m_d$) is in the unit of keV/$c^2$ and $\sigma_{pe}$ in the unit of barn.
The constants $g_{ae}$ and $\kappa$ are measured in experiments.
Among the searches of ALPs and DPs~\cite{XENON100:2017pfn,XENON:2022ltv,GERDA:2024gip,COSINE-100:2023dir,XMASS:2018pvs,LUX:2017glr,PandaX:2017ock,EDELWEISS:2018tde,Majorana:2022gtu,LZ:2023poo}, XENONnT~\cite{XENON:2022ltv} has the leading limit at the masses below 140~keV/$c^2$, while GERDA~\cite{GERDA:2024gip} and COSINE-100~\cite{COSINE-100:2023dir} have set the most significant constraints in the
$\mathcal{O}$(100)~keV/$c^2$ to 1~MeV/$c^2$ range.
Note that a strong cosmological constraint on axionlike particles (ALPs) with masses above 100 keV/$c^2$ can be derived from the lifetime requirement, assuming that heavy ALPs couple exclusively to electrons or have generic couplings to other fermions~\cite{Ferreira:2022egk}. However, if these assumptions about the couplings are relaxed, the resulting constraints carry significant uncertainties. 
Consequently, the direct search for ALPs in this mass range remains a topic of high scientific interest~\cite{Ferreira:2022egk}.

In this Letter, we use the commissioning dataset (Run0) and the first scientific dataset (Run1) of the PandaX-4T experiment, covering the periods from November 28, 2020, to April 16, 2021, and November 16, 2021, to May 15, 2022, respectively, to search for ALP and DP signals. The durations of Run0 and Run1 are 94.8 day and 163.5 day, respectively.
The targeted ALP or DP masses are between 30~keV/$c^2$ and 1~MeV/$c^2$, while the search is performed in an energy region of interest~(ROI) of 25 to 1050~keV.  
Compared to previous analyses of PandaX-4T in the MeV energy range~\cite{PandaX:2022kwg, PandaX:2023ggs}, the energy reconstruction procedure is further optimized to improve the energy resolution.
The time-varying background contributions from short-lived xenon isotopes, including $^{127}$Xe, $^\textrm{129m}$Xe, and $^\textrm{131m}$Xe, are now incorporated into the modeling for the first time in PandaX-4T.
Furthermore, we have developed a convolution method to propagate uncertainties of energy response into the energy spectrum to fully incorporate detector uncertainties in the likelihood fit.

The PandaX-4T detector is a cylindrical, dual-phase time projection chamber~(TPC) measuring 118.5~cm in diameter and 118.5~cm in height. 
The active volume containing 3.7 ton of natural xenon is surrounded by a field cage with an anode on the top and a cathode on the bottom. 
Two three-inch Hamamatsu PMT arrays are installed above the anode and below the cathode for signal readout. 
A detailed description of the detector can be found in Ref.~\cite{PandaX-4T:2021bab}.
The detector measures the energy deposition and its three-dimensional position via the scintillation signal ($S1$) and the electroluminescence signal ($S2$), which scales with number of ionized electrons.

The data production and event selection procedures are similar to Refs.~\cite{PandaX:2022kwg, PandaX:2023ggs,Lu:2024ilt,PandaX:2024med}, where the reconstruction of single-site~(SS) spectrum from 25~keV to 2.8~MeV is achieved.
To avoid systematic effects due to saturation of the top PMTs, the total energy of an event is calculated by combining the charges of $S1$ and $S2_\mathrm{b}$, collected by the bottom PMT array, according to the formula $E = 13.7~\mathrm{eV} \times (S1/g_1+S2_\mathrm{b}/g_{2_\mathrm{b}})$~\cite{Szydagis:2011tk}.
The detector parameters ($g_1$, $g_{2_\mathrm{b}}$) are prefitted by the mono-energy electronic recoil peaks~\cite{PandaX:2024xpq}.

The horizontal position is obtained based on the maximum likelihood estimation with the desaturated charge pattern of $S2$ in the top PMT array and the photon acceptance functions derived from optical Monte Carlo (MC) simulations.
We have optimized the position reconstruction in the vertical ($z$), radial ($R$) and azimuthal ($\phi$) directions using calibration data from $^\textrm{83m}$Kr and wall events from $^{210}$Po $\alpha$ particles. 
The same fiducial volume cuts of the previous study~\cite{PandaX:2023ggs} have been adopted in this analysis.
The fiducial mass~(FM) is determined to be $625\pm10$~kg for Run0 and $621\pm13$~kg for Run1.
These values are obtained by scaling the percentage of \(^{83\text{m}}\mathrm{Kr}\), corresponding to a total exposure of 440~kg$\cdot$yr. 
The uncertainty is determined by the LXe density, and by the difference between the geometrically calculated and \(^{83\text{m}}\mathrm{Kr}\) rate-scaled volumes~\cite{PandaX:2023ggs}.

\begin{table}[tbp]
   \caption{
   Summary of sources of systematic uncertainties. 
   $\mathcal{M}_0$ denotes the five-parameter detector response model (see text), with its means and uncertainties determined from mono-energetic peaks obtained during calibration runs and outside the ROI.
   }
   \label{tab:sys_err}
   \centering
   \renewcommand{\arraystretch}{1.5}
    \resizebox{\linewidth}{!}{
   \begin{tabular}{>{\centering\arraybackslash}m{.5cm}>{\centering\arraybackslash}m{.5cm}>{\centering\arraybackslash}m{.5cm}>{\centering\arraybackslash}m{.5cm}}
       \toprule
       \multicolumn{2}{c}{Sources} & \multicolumn{1}{c}{Run0} & \multicolumn{1}{c}{Run1} \\ 
       \hline
       \multicolumn{1}{c}{\multirow{5}{*}{\shortstack[t]{Detector \\ response}}}&\multicolumn{1}{c}{\multirow{1}{*}{\shortstack[t]{$a_0$~[$\sqrt{\mathrm{keV}}$]}}} & \multicolumn{1}{c}{$0.43\pm0.02$} & \multicolumn{1}{c}{$0.45\pm0.02$} \\
       \multicolumn{1}{c}{\multirow{2}{*}{}}&\multicolumn{1}{c}{\multirow{1}{*}{\shortstack[t]{$b_0$~[keV$^{-1}$]}}} & \multicolumn{1}{c}{$(5\pm2) \times 10^{-6}$} & \multicolumn{1}{c}{$(5\pm2) \times 10^{-6}$} \\
       \multicolumn{1}{c}{\multirow{2}{*}{}}&\multicolumn{1}{c}{\multirow{1}{*}{\shortstack[t]{$c_0$}}} & \multicolumn{1}{c}{$(-7\pm20) \times 10^{-4}$} & \multicolumn{1}{c}{$(-7\pm22) \times 10^{-4}$} \\
       \cline{2-4}
       \multicolumn{1}{c}{\multirow{2}{*}{}}&\multicolumn{1}{c}{\multirow{1}{*}{\shortstack[t]{$d_0$}}} & \multicolumn{1}{c}{$0.9930\pm0.0008$} & \multicolumn{1}{c}{$0.9989\pm0.0009$} \\
       \multicolumn{1}{c}{\multirow{2}{*}{}}&\multicolumn{1}{c}{\multirow{1}{*}{\shortstack[t]{$e_0$~[keV]}}} & \multicolumn{1}{c}{$0.74\pm0.06$} & \multicolumn{1}{c}{$1.25\pm0.06$} \\
       \cline{1-4}
       \multicolumn{1}{c}{\multirow{2}{*}{\shortstack[t]{Overall \\ efficiency}}}&\multicolumn{1}{c}{SS fraction (1 MeV/$c^2$)} & \multicolumn{1}{c}{($96 \pm 4$)\%} & \multicolumn{1}{c}{($96 \pm 4$)\%} \\
       \multicolumn{1}{c}{\multirow{2}{*}{}}&\multicolumn{1}{c}{Quality cut} & \multicolumn{1}{c}{($99.87\pm0.02$)\%} & \multicolumn{1}{c}{($99.75\pm0.10$)\%} \\
       \cline{1-4}
        \multicolumn{1}{c}{\multirow{2}{*}{\shortstack[t]{Signal \\ selection}}}&\multicolumn{1}{c}{LXe density~[g/cm$^{3}$]} & \multicolumn{2}{c}{$2.850\pm0.004$} \\
       \multicolumn{1}{c}{\multirow{2}{*}{}}&\multicolumn{1}{c}{FV uniformity~[kg]} & \multicolumn{1}{c}{$625\pm10$} & \multicolumn{1}{c}{$621\pm13$} \\
       \cline{1-4}
       \multicolumn{2}{c}{Background model} & \multicolumn{2}{c}{Table~\ref{tab:bkg_summary}} \\
       \toprule
   \end{tabular}
   }
\end{table}

The detector is calibrated using multiple mono-energetic peaks observed in the data, obtained either during dedicated calibration periods or physics runs. 
The nonuniformity in the spatial energy response is monitored and corrected using the uniformly distributed 41.5 keV peak from $^\textrm{83m}$Kr injection data. 
Compared to the previous analysis~\cite{PandaX:2023ggs}, the energy reconstruction has been improved in two key aspects.
First, the temporal variations in light yield and charge yield are characterized using the $\alpha$ signals from $^{222}$Rn progenies and corrected accordingly. 
This correction has led to an improvement in energy resolution, e.g., from 3.6\% to 3.0\% at 208 keV in Run0, enabling more accurate estimation of the activities of short-lived xenon isotopes (see later). 
Second, the energy response model has been further refined by incorporating the $^\textrm{83m}$Kr peak, along with the 164 keV peak (from $^\textrm{131m}$Xe), the 236 keV peaks (from $^\textrm{127}$Xe and $^\textrm{129m}$Xe) obtained during calibration runs, and the 1460 keV peak (from $^{40}$K) outside the ROI. 
These calibration data points are completely uncorrelated with those used in the final spectral fit. 
The inclusion of the $^\textrm{83m}$Kr peak allows the analysis window to be extended down to 25 keV. 
The energy response is modeled using five parameters, independently for Run0 and Run1. 
The energy resolution is modeled as a Gaussian function with the width $\sigma(E)$ constructed as $\frac{\sigma(E)}{E}=\frac{a}{\sqrt{E}} + b \cdot E + c$, where $E$ is the reconstructed energy in the unit of keV.
The residual energy nonlinearity is modeled as $E = d \cdot \hat{E} + e$ where $\hat{E}$ is the true energy. 
The calibrated values and uncertainties $\mathcal{M}_0={(a_0, b_0, c_0, d_0, e_0)^\textrm{T}}$ (Table~\ref{tab:sys_err}) and the $5 \times 5$ covariance matrix $\Sigma_m$ will be used in fitting the Run0 and Run1 data.

The total detection efficiencies for signals of ALPs and DPs are the product of SS cut efficiency, data quality cut efficiency, and ROI acceptance.
The identification of SS and multi-site (MS) events follows the method in Ref.~\cite{PandaX:2022kwg}. 
Charge deposits in a given event may be separated into different $S2$ clusters, allowing classification as SS or MS based on the number of observed $S2$ peaks. 
The ratio of SS to SS+MS events within the ROI is calculated using BambooMC, a GEANT4-based Monte Carlo (MC) framework~\cite{Chen:2021asx}, with the LXe response to ER modeled using NEST 2.0~\cite{szydagis2011nest}, and validated through $^{232}$Th calibration data.
The SS fraction for the signal varies ranging from 100.0\% to 95.7\% within the ROI.
The relative systematic uncertainty from $^{232}$Th calibration data is conservatively applied on both the signal and background for the SS fraction, which is obtained from the difference between data and simulation, averaged over the energy range.
Quality cut variables, used to eliminate noise and select electronic recoil events, are adopted from Ref.~\cite{PandaX:2023ggs} but have been adjusted to account for events down to 25 keV. 
The adjusted cut criteria are validated using calibration data and subsequently applied to the entire Run0 and Run1, resulting in an efficiency of ($99.87\pm0.02$)\% and ($99.75\pm0.10$)\%, respectively.
The ROI acceptance for signal is close to 100\%.

In our ROI, the background contribution originates from the detector materials, liquid xenon, and solar neutrinos, as shown in Table~\ref{tab:bkg_summary}.
The activities of $^{232}$Th, $^{238}$U, $^{60}$Co, and $^{40}$K in detector materials have been reported in Ref.~\cite{PandaX:2022kwg}.
The concentration of $^{\textrm{85}}$Kr is determined by $\beta$-$\gamma$ cascades through the metastable state $^{\textrm{85m}}$Rb. 
The Kr/Xe concentration in Run0 (Run1) is \(0.52\pm0.27\) (\(0.94\pm0.28\)) parts per trillion~\cite{PandaX:2024qfu}, assuming an isotopic abundance of \(2 \times 10^{-11}\) for $^{\textrm{85}}$Kr~\cite{Collon:2004xs}.
The $^{214}$Pb rate is left float in the fit, the same as the previous analyses~\cite{PandaX:2023ggs}. The $^{212}$Pb originated from $^{220}$Rn emanation is a subdominant background component. Because of the strong dependence of $^{212}$Pb to the circulation conditions, etc., the rate of $^{212}$Pb of each run in the fit is also set free.

The energy spectrum of the elastic scattering of solar $pp$ and $^7$Be neutrinos on electrons is adopted from the Ref.~\cite{Chen:2016eab}, with an uncertainty of approximately 10\%~\cite{BOREXINO:2014pcl}, resulting in $82\pm9$ ($140\pm15$) events within ROI in Run0 (Run1).

 \begin{figure}[tb]
    \includegraphics[width=1.\columnwidth]{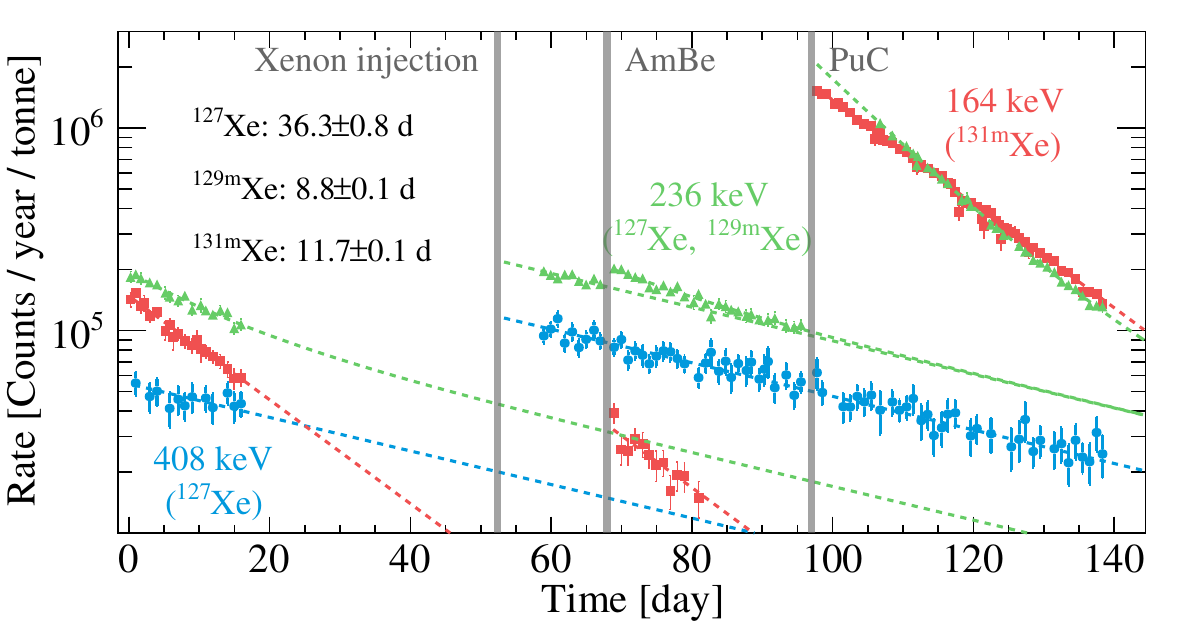}
    \caption{Time evolution of $^{127}$Xe, $^\textrm{129m}$Xe and $^\textrm{131m}$Xe in the SS spectrum during the Run0 of PandaX-4T experiment. 
    The starting time of the horizontal axis is defined as the beginning of the dataset.
    AmBe and PuC refer to sources used in the neutron calibrations.
 }
    \label{fig:xe_evolution}
\end{figure}

\begin{table}[t]
    \caption{
    The total background contributions in the ROI are presented for both Run0 and Run1. 
    The fitted counts are obtained from the background-only fit.}
    \label{tab:bkg_summary}
    \centering
    \renewcommand{\arraystretch}{1.3}
    \begin{tabular}{>{\centering\arraybackslash}m{2.5cm}>{\centering\arraybackslash}m{2.5cm}>{\centering\arraybackslash}m{2.5cm}}
    \toprule
    \multicolumn{1}{c}{Components} & \multicolumn{1}{c}{Expected ($\times 10^2$)} & \multicolumn{1}{c}{Fitted ($\times 10^2$)} \\
    \midrule
    $^{232}$Th & $9.7 \pm 5.8$ & $12.7 \pm 2.5$ \\
    $^{238}$U & $3.8 \pm 2.8$ & $6.6 \pm 2.2$ \\
    $^{60}$Co & $5.8 \pm 3.5$ & $9.3 \pm 3.0$ \\
    $^{40}$K & $4.5 \pm 2.0$ & $5.6 \pm 1.9$ \\
    \midrule
    $^{85}$Kr & $19.0 \pm 4.9$ & $26.4 \pm 2.3$ \\
    $^{214}$Pb & float & $352.9 \pm 7.5$  \\
    $^{212}$Pb & float & $18.6 \pm 2.5$ \\
    $^{136}$Xe & $352 \pm 16$ & $358.5 \pm 9.1$ \\ 
    $^{124}$Xe & $1.37 \pm 0.21$ & $1.41 \pm 0.13$ \\
    $^{125}$Xe (Run0) & float & $6.48 \pm 0.83$ \\
    $^{125}$I & $0.66 \pm 0.16$ & $0.59 \pm 0.13$\\
    $^{133}$Xe (Run0) & float & $86.1 \pm 2.2$ \\
    164~keV (Run0) & $414 \pm 17$ & $407.7 \pm 6.4$ \\
    164~keV (Run1) & float & $4.67 \pm 0.32$ \\  
    208~keV (Run0) & $37.8 \pm 1.3$ & $37.88 \pm 0.77$ \\
    236~keV (Run0) & $565 \pm 66$ & $560.8 \pm 8.8$ \\
    236~keV (Run1) & float & $3.05 \pm 0.32$ \\
    380~keV (Run0) & $24.3 \pm 1.2$ & $24.10 \pm 0.66$ \\
    408~keV (Run0)  & $87.9 \pm 3.2$ & $88.8 \pm 1.6$ \\
    \midrule
    $pp+^{7}$Be $\nu$ & $2.22 \pm 0.24$ & $2.31 \pm 0.23$ \\
     \bottomrule
    \end{tabular}
\end{table}

\begin{figure*}[tb]
    \includegraphics[width=1.\textwidth]{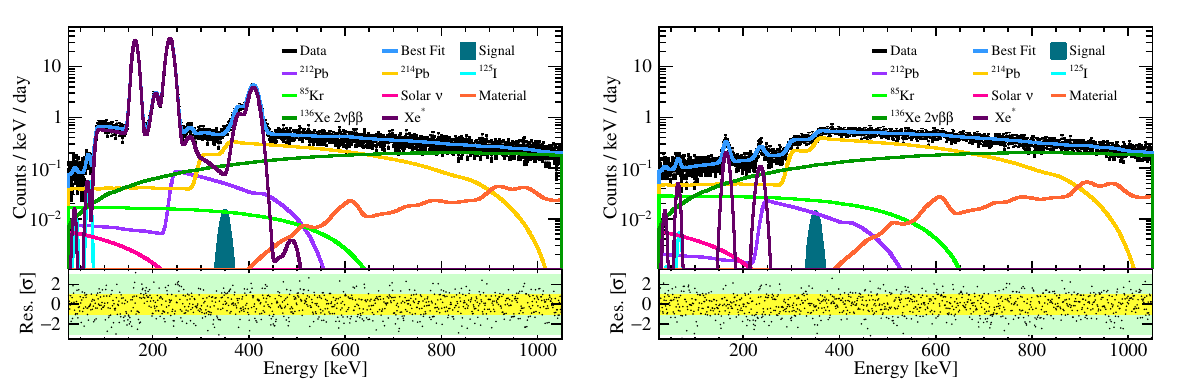}
    \caption{
    The background-only SS data and fit, along with hypothetical 350 keV/$c^2$ DM signals (ALP with $g_{ae} = 1.0 \times 10^{-12}$, DP with $\kappa = 2.0 \times 10^{-12}$), are shown for Run0 (left) and Run1 (right) from 25 to 1050 keV, with a bin size of 1 keV.
    The horizontal axis represents the reconstructed energy in the data.
    Xe$^*$ includes the contributions from $^{124}$Xe, $^{125}$Xe, $^{127}$Xe, $^\textrm{129m}$Xe, $^\textrm{131m}$Xe, and $^{133}$Xe.
    The lower panel shows the residuals together with $\pm1\sigma$~($\pm3\sigma$) bands.}
    \label{fig:fit_example}
\end{figure*}

The contributions from $^{136}$Xe two-neutrino double beta decay ($2\nu\beta\beta$) and $^{124}$Xe double electron capture are calculated based on the half-life measurements reported in PandaX-4T~\cite{PandaX:2022kwg, PandaX:2024xpq}. 
The expected events for $^{136}$Xe $2\nu\beta\beta$ within the ROI are $13000\pm598$ for Run0 and $22211\pm1,022$ for Run1, respectively. 
Similarly, the expected events for $^{124}$Xe double electron capture within the ROI are $51\pm8$ for Run0 and $86\pm13$ for Run1.

Short-lived xenon isotopes, $^\textrm{125}$Xe, $^\textrm{129m}$Xe, $^\textrm{131m}$Xe and $^\textrm{133}$Xe were induced by neutron calibration.
The neutron sources, including $^\textrm{241}$Am-Be and $^\textrm{238}$Pu-C, are used before, during, and after Run0, while no neutron calibration runs are performed during Run1.
$^\textrm{129m}$Xe and $^\textrm{131m}$Xe results in mono-energetic peaks at 236 and 164~keV, with the half-life of 8.9 and 11.8~day, respectively. 
$^{125}$Xe, with a half-life of 16.9 h, undergoes electron capture and decays to the relatively long-lived isotope $^{125}$I, with a half-life of 59.4~day.
The spectrum of $^{133}$Xe is continuous, featuring a $\beta$ spectrum with a 346 keV endpoint in combination with a 81 keV deexcitation $\gamma$ ray.
In addition, $^{127}$Xe was introduced by a batch of approximately 30~kg of xenon from above-ground (exposed to cosmogenic neutrons), injected into the detector during the Run0 data taking.
$^{127}$Xe decays via electron capture with a half-life of 36.4 day. 
The deexcitation of the daughter $^{127}$I generates mono-energetic peaks at 208, 236, 380, and 408~keV. 

The background contributions from these neutron-activated components are summarized Table~\ref{tab:bkg_summary}. 
Because of their clear time dependences, the expected activities of $^{127}$Xe, $^\textrm{129m}$Xe, and $^\textrm{131m}$Xe in Run0 are prefitted according to the time evolution in the SS data, with decay half-lives set free.
In order to increase statistics, a larger FM of 2.43~ton is selected.
To get the rate evolution, in every time division, each peak in the energy spectrum is locally characterized by a Gaussian function plus a linear background.
The evolution of $^\textrm{131m}$Xe is fitted as a single component, while the evolution of four peaks of $^\textrm{127}$Xe and $^\textrm{129m}$Xe is fitted simultaneously to decompose the contribution of two isotopes at 236~keV.
The resulting time evolution and the corresponding measured half-lives are shown in Fig.~\ref{fig:xe_evolution}, with fitted half-lives consistent with existing values in nuclear databases~\cite{nndc_nudat}.
On the other hand, the activities of $^{125}$Xe and $^{133}$Xe are determined later by the spectral likelihood fit. 
For Run1, the level of $^{125}$Xe, $^{133}$Xe, and $^{127}$Xe have decayed to zero. 
Small amounts of $^\textrm{129m}$Xe and $^\textrm{131m}$Xe in Run1 do not allow a meaningful prefit either, therefore are left float in later spectral fit. 
For the activity of $^{125}$I in both runs,  we adopt the results from a two-component exponential model fit in Ref.~\cite{PandaX:2024xpq}.

The final SS energy spectral fit is then performed using a one-dimensional binned likelihood function constructed as 
\begin{equation}
\begin{aligned}
 L = & \prod_{r=0}^{1} \prod_{i=1}^{N_\mathrm{bins}} \frac{(N_{r,i})^{N_{r,i}^\mathrm{obs}}e^{-N_{r,i}}}{N_{r,i}^\mathrm{obs}!}
 \mathcal{G}(\mathcal{M}_r; \mathcal{M}_r^0, \Sigma_r) \\
 & \cdot \prod_{j=1}^{N_\mathrm{G}} G(\eta_j; 0, \sigma_j)
 ,
\end{aligned}
\label{eq::likelihood}
\end{equation}
where $N_{r,i}$ and $N_{r,i}^\mathrm{obs}$ are the expected and observed events numbers of the $i_{th}$ energy bin in the Run-$r$, respectively. 
$N_i$ in Run-$r$ is defined as
\begin{equation}
\begin{aligned}
 N_{i} = & (1+\eta_{a}) \cdot [ (1+\eta_s) \cdot n_{s} \cdot S_{i} + \sum_{b=1}^{N_\mathrm{bkg}} (1+\eta_b) \cdot n_{b} \cdot B_{b, i}],
\end{aligned}
\end{equation}
where $n_s$ and $n_b$ are the counts of signal $s$ and background component $b$, respectively.
The corresponding $S_i$ and $B_{b,i}$ are the $i_\mathrm{th}$ bin values of the normalized energy spectrum convolved with the five-parameter energy response model.
The Gaussian penalty term $\mathcal{G}(\mathcal{M}_r; \mathcal{M}_r^0, \Sigma_r)$ of the energy response contains the five-parameter $\mathcal{M}_r^0$ (Table~\ref{tab:sys_err}) and the covariant matrix $\Sigma_r$ in Run-$r$.
The Gaussian penalty terms $G(\eta_j; 0, \sigma_j)$ are used to constrain the nuisance parameters $\eta_{a}$, $\eta_{s}$, and $\eta_{b}$, which represent the relative uncertainties in the overall efficiency (Table~\ref{tab:sys_err}), the signal selection (Table~\ref{tab:sys_err}), and background model (Table~\ref{tab:bkg_summary}), respectively.
The activities of $^{232}$Th, $^{238}$U, $^{60}$Co, $^{40}$K, $^{124}$Xe, $^{136}$Xe, as well as the solar $pp$ and $^{7}$Be fluxes are identical in Run0 and Run1. 
Other background components are treated independently for Run0 and Run1.

A background-only fit is performed prior to the signal fits, yielding a $\chi^2$/NDF of 1.06, as illustrated in Fig.~\ref{fig:fit_example}.
The data are consistent with the background-only model, with a $p$-value of 0.51.
The contributions of other fitted background components, summarized in Table~\ref{tab:bkg_summary}, are consistent with their expected values, except $^{\textrm{85}}$Kr which is pulled slightly upward by 1.5 $\sigma$.
Background is dominated by short-lived xenon isotopes, $^{136}$Xe 2$\nu\beta\beta$, $^{214}$Pb $\beta$ decay, and detector material.
The detector response nuisance parameters ($5 \times 2$) for the two runs are within 1~$\sigma$ of their input values, except for parameter $a$ and $c$ in the energy response model of Run0 and $e$ in that of Run1, which are pulled by 2.9~$\sigma$, 1.8~$\sigma$, and 1.2~$\sigma$, respectively. 
This suggests that the energy resolution function derived solely from mono-energetic peaks is insufficient to describe the full measured spectrum. 
However, we have verified that these deviations have a negligible impact on the limits of the couplings of ALPs and DPs.

We conduct a scanned fit to the SS spectrum, including Gaussian peaks of the hypothetical DM signals.
The DM masses range from 30~keV/$c^2$ to 1~MeV/$c^2$ with a step size of 10~keV/$c^2$.
The local significances of five DM masses are found to range between 2~$\sigma$ and 3~$\sigma$.
For instance, the significance at 230 keV, which occurrs near the Gaussian background of 236~keV~(from $^\textrm{127}$Xe and $^\textrm{129m}$Xe), is 3.0~$\sigma$.
Taking into account the look elsewhere effect~\cite{Ranucci:2012ed, Cowan:2010js}, the global significance only reaches 1.5~$\sigma$. 
Therefore, no significant evidence for a signal is observed within the mass range of [30~keV/$c^2$, 1~MeV/$c^2$]. 

\begin{figure}[tbp]
    \centering
    \includegraphics[width=1.\columnwidth]{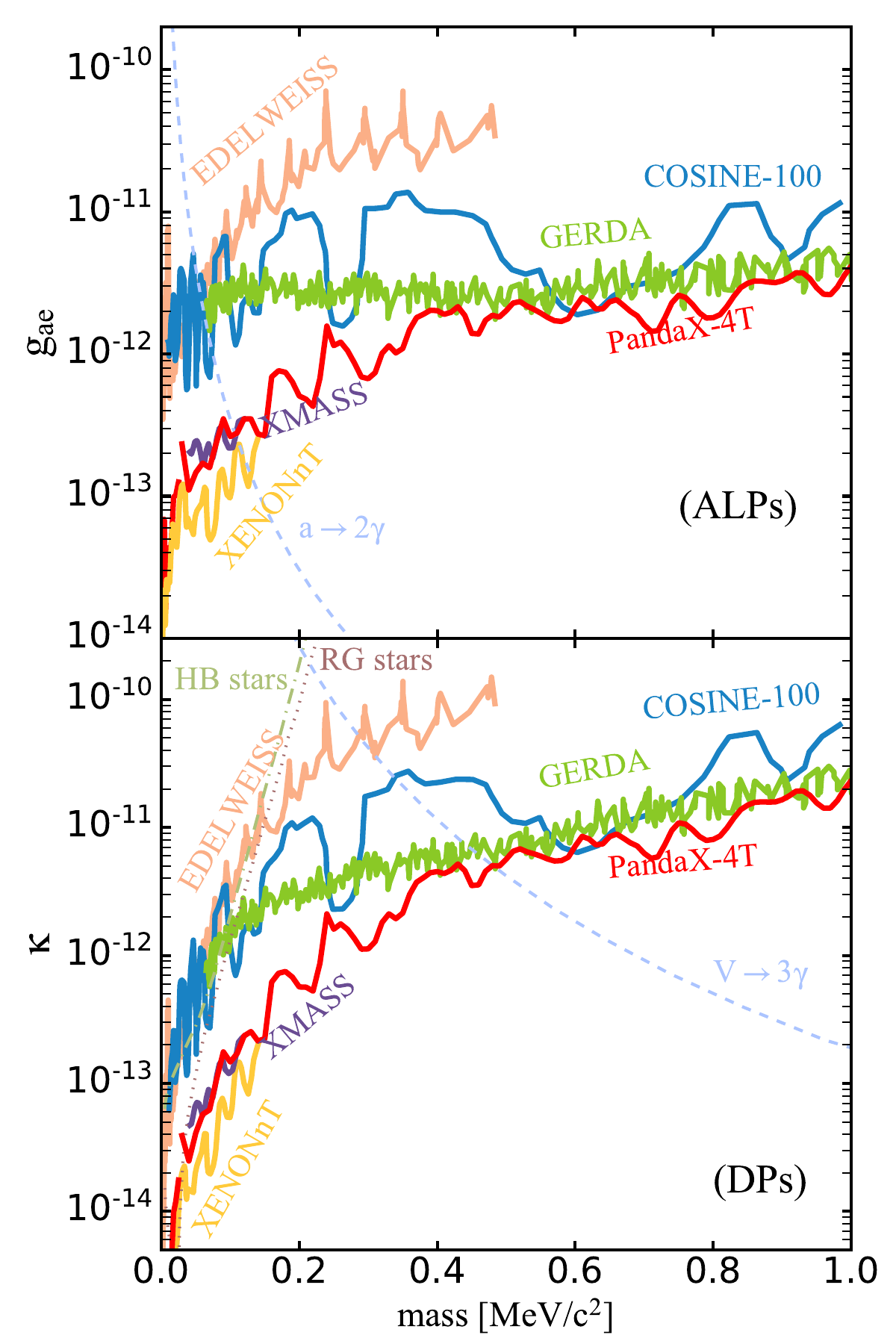}
    \caption{
       The 90\% C.L. upper limits of the couplings of ALPs (top) and DPs (bottom) to atomic electrons. 
       The results from this work are represented by the red line starting at 30 keV, while the line below 25 keV is taken from Ref.~\cite{PandaX:2024cic}.
       The dashed blue lines indicates the stability requirement for ALPs and DPs over the age of the Universe\cite{Ferreira:2022egk,Redondo:2008ec}.
       The dotted brown line and the dot-dashed green line represent constraints derived from energy losses in red giant (RG) and horizontal branch (HB) stars, respectively~\cite{Li:2023vpv}.
    }
    \label{fig:exclusion_curve}
\end{figure}

The upper limits at 90\% confidence level~(C.L.) on the event rate have been set and converted to upper limits of coupling strength (Fig.~\ref{fig:exclusion_curve}), as described in Eq.~(\ref{eq:event_rate}).
Leading direct detection limits from other experiments are also plotted for comparison~\cite{EDELWEISS:2018tde,XENON:2022ltv,COSINE-100:2023dir,GERDA:2024gip}. 
The relative deterioration at certain masses in the limit curves is due to background fluctuations, such as the mono-energetic background at 164~keV from $^\textrm{131m}$Xe.
Our limits are the most competitive over a broad mass range, spanning from 150~keV/$c^2$ to 1~MeV/$c^2$, with an average improvement with a factor of 2.0 compared to existing results~\cite{COSINE-100:2023dir,GERDA:2024gip}.
Notably, the improvement is particularly significant in the \text{150-400} keV/$c^2$ range, where the enhancement factor reaches 3.5.
The largest improvement (a factor of 13.4) occurs around 150~keV/$c^2$, constraining the coupling of ALPs (DPs) to electrons to $g_{ae}<2.7 \times 10^{-13}$ ($\kappa < 2.3\times 10^{-13}$).
Compared to other experiments, the improvement is due to a combination of large exposure, low background rate, and broader energy range.
It is noteworthy that the weakening of our upper limits with increasing mass is primarily due to the steep decrease of the photoelectric cross section, as we solely searched for absorption events.
In the MeV mass regions of the ALPs and DPs, the cross section of Compton-like process~\cite{COSINE-100:2023dir}, which are mostly MS events, becomes more prominent. 
A dedicated analysis focused on MS events is underway to improve the signal detection efficiency.

In summary, we have searched for ALPs and DPs with masses up to 1 MeV/$c^2$ using 440 kg$\cdot$yr exposure of PandaX-4T Run0 and Run1. 
A detailed analysis of the time evolution of xenon isotopes improves the background modeling, and the inclusion of energy response model convolution in the likelihood function results in a more rigorous treatment of systematic uncertainties.
No significant excess over the expected background is observed, and the upper limits at 90\% C.L. on the effective couplings between ALPs/DPs and atomic electrons of xenon are derived.
Our limit is comparable with other direct searches in the DM mass range from 30 to 150~keV/$c^2$ and the most competitive over a large mass range from 150~keV/$c^2$ to 1~MeV/$c^2$. 
Between 2023 and 2024, PandaX-4T underwent upgrades to its PMTs, electronics, DAQ, and external water veto systems. 
The second science run (Run2) is now underway, accompanied by continuous efforts to further reduce background levels.
Larger statistics and higher-quality data can further enhance the understanding of the background model and the search sensitivities of the  ALPs and DPs. 
% !TEX root = ../main.

%\section{Acknowledgement}

This project is supported in part by grants from the Ministry of Science and Technology of China (No. 2023YFA1606200, No. 2023YFA1606201, No. 2023YFA1606202), National Science Foundation of China (No. 12090060, No. 12090061, No. 12305121, No. U23B2070), China Postdoctoral Science Foundation (No. 2023M744093), and Office of Science and Technology, Shanghai Municipal Government (Grants No. 22JC1410100, No. 21TQ1400218).
We are grateful for the support by the Fundamental Research Funds for the Central Universities. We also are thankful for the sponsorship from Hongwen Foundation in Hong Kong, New Cornerstone Science Foundation, Tencent Foundation in China, and Yangyang Development Fund. 
Finally, we thank the CJPL administration and the Yalong River Hydropower Development Company Ltd. for indispensable logistical support and other help. 
\bibliographystyle{apsrev4-1}
\bibliography{P4SuperWIMPs}

%merlin.mbs apsrev4-1.bst 2010-07-25 4.21a (PWD, AO, DPC) hacked
%Control: key (0)
%Control: author (72) initials jnrlst
%Control: editor formatted (1) identically to author
%Control: production of article title (-1) disabled
%Control: page (0) single
%Control: year (1) truncated
%Control: production of eprint (0) enabled
\begin{thebibliography}{48}%
\makeatletter
\providecommand \@ifxundefined [1]{%
 \@ifx{#1\undefined}
}%
\providecommand \@ifnum [1]{%
 \ifnum #1\expandafter \@firstoftwo
 \else \expandafter \@secondoftwo
 \fi
}%
\providecommand \@ifx [1]{%
 \ifx #1\expandafter \@firstoftwo
 \else \expandafter \@secondoftwo
 \fi
}%
\providecommand \natexlab [1]{#1}%
\providecommand \enquote  [1]{``#1''}%
\providecommand \bibnamefont  [1]{#1}%
\providecommand \bibfnamefont [1]{#1}%
\providecommand \citenamefont [1]{#1}%
\providecommand \href@noop [0]{\@secondoftwo}%
\providecommand \href [0]{\begingroup \@sanitize@url \@href}%
\providecommand \@href[1]{\@@startlink{#1}\@@href}%
\providecommand \@@href[1]{\endgroup#1\@@endlink}%
\providecommand \@sanitize@url [0]{\catcode `\\12\catcode `\$12\catcode
  `\&12\catcode `\#12\catcode `\^12\catcode `\_12\catcode `\%12\relax}%
\providecommand \@@startlink[1]{}%
\providecommand \@@endlink[0]{}%
\providecommand \url  [0]{\begingroup\@sanitize@url \@url }%
\providecommand \@url [1]{\endgroup\@href {#1}{\urlprefix }}%
\providecommand \urlprefix  [0]{URL }%
\providecommand \Eprint [0]{\href }%
\providecommand \doibase [0]{http://dx.doi.org/}%
\providecommand \selectlanguage [0]{\@gobble}%
\providecommand \bibinfo  [0]{\@secondoftwo}%
\providecommand \bibfield  [0]{\@secondoftwo}%
\providecommand \translation [1]{[#1]}%
\providecommand \BibitemOpen [0]{}%
\providecommand \bibitemStop [0]{}%
\providecommand \bibitemNoStop [0]{.\EOS\space}%
\providecommand \EOS [0]{\spacefactor3000\relax}%
\providecommand \BibitemShut  [1]{\csname bibitem#1\endcsname}%
\let\auto@bib@innerbib\@empty
%</preamble>
\bibitem [{\citenamefont {Aghanim}\ \emph {et~al.}(2020)\citenamefont {Aghanim}
  \emph {et~al.}}]{Planck:2018vyg}%
  \BibitemOpen
  \bibfield  {author} {\bibinfo {author} {\bibfnamefont {N.}~\bibnamefont
  {Aghanim}} \emph {et~al.} (\bibinfo {collaboration} {Planck}),\ }\href
  {\doibase 10.1051/0004-6361/201833910} {\bibfield  {journal} {\bibinfo
  {journal} {Astron. Astrophys.}\ }\textbf {\bibinfo {volume} {641}},\ \bibinfo
  {pages} {A6} (\bibinfo {year} {2020})},\ \bibinfo {note} {[Erratum:
  Astron.Astrophys. 652, C4 (2021)]},\ \Eprint
  {http://arxiv.org/abs/1807.06209} {arXiv:1807.06209 [astro-ph.CO]}
  \BibitemShut {NoStop}%
\bibitem [{\citenamefont {Bergstrom}(2012)}]{Bergstrom:2012fi}%
  \BibitemOpen
  \bibfield  {author} {\bibinfo {author} {\bibfnamefont {L.}~\bibnamefont
  {Bergstrom}},\ }\href {\doibase 10.1002/andp.201200116} {\bibfield  {journal}
  {\bibinfo  {journal} {Annalen Phys.}\ }\textbf {\bibinfo {volume} {524}},\
  \bibinfo {pages} {479} (\bibinfo {year} {2012})},\ \Eprint
  {http://arxiv.org/abs/1205.4882} {arXiv:1205.4882 [astro-ph.HE]} \BibitemShut
  {NoStop}%
\bibitem [{\citenamefont {Feng}(2010)}]{Feng:2010gw}%
  \BibitemOpen
  \bibfield  {author} {\bibinfo {author} {\bibfnamefont {J.~L.}\ \bibnamefont
  {Feng}},\ }\href {\doibase 10.1146/annurev-astro-082708-101659} {\bibfield
  {journal} {\bibinfo  {journal} {Ann. Rev. Astron. Astrophys.}\ }\textbf
  {\bibinfo {volume} {48}},\ \bibinfo {pages} {495} (\bibinfo {year} {2010})},\
  \Eprint {http://arxiv.org/abs/1003.0904} {arXiv:1003.0904 [astro-ph.CO]}
  \BibitemShut {NoStop}%
\bibitem [{\citenamefont {Lee}\ and\ \citenamefont
  {Weinberg}(1977)}]{Lee:1977ua}%
  \BibitemOpen
  \bibfield  {author} {\bibinfo {author} {\bibfnamefont {B.~W.}\ \bibnamefont
  {Lee}}\ and\ \bibinfo {author} {\bibfnamefont {S.}~\bibnamefont {Weinberg}},\
  }\href {\doibase 10.1103/PhysRevLett.39.165} {\bibfield  {journal} {\bibinfo
  {journal} {Phys. Rev. Lett.}\ }\textbf {\bibinfo {volume} {39}},\ \bibinfo
  {pages} {165} (\bibinfo {year} {1977})}\BibitemShut {NoStop}%
\bibitem [{\citenamefont {Goodman}\ and\ \citenamefont
  {Witten}(1985)}]{Goodman:1984dc}%
  \BibitemOpen
  \bibfield  {author} {\bibinfo {author} {\bibfnamefont {M.~W.}\ \bibnamefont
  {Goodman}}\ and\ \bibinfo {author} {\bibfnamefont {E.}~\bibnamefont
  {Witten}},\ }\href {\doibase 10.1103/PhysRevD.31.3059} {\bibfield  {journal}
  {\bibinfo  {journal} {Phys. Rev. D}\ }\textbf {\bibinfo {volume} {31}},\
  \bibinfo {pages} {3059} (\bibinfo {year} {1985})}\BibitemShut {NoStop}%
\bibitem [{\citenamefont {Bertone}\ \emph {et~al.}(2005)\citenamefont
  {Bertone}, \citenamefont {Hooper},\ and\ \citenamefont
  {Silk}}]{Bertone:2004pz}%
  \BibitemOpen
  \bibfield  {author} {\bibinfo {author} {\bibfnamefont {G.}~\bibnamefont
  {Bertone}}, \bibinfo {author} {\bibfnamefont {D.}~\bibnamefont {Hooper}}, \
  and\ \bibinfo {author} {\bibfnamefont {J.}~\bibnamefont {Silk}},\ }\href
  {\doibase 10.1016/j.physrep.2004.08.031} {\bibfield  {journal} {\bibinfo
  {journal} {Phys. Rept.}\ }\textbf {\bibinfo {volume} {405}},\ \bibinfo
  {pages} {279} (\bibinfo {year} {2005})},\ \Eprint
  {http://arxiv.org/abs/hep-ph/0404175} {arXiv:hep-ph/0404175} \BibitemShut
  {NoStop}%
\bibitem [{\citenamefont {Liu}\ \emph {et~al.}(2017)\citenamefont {Liu},
  \citenamefont {Chen},\ and\ \citenamefont {Ji}}]{Liu:2017drf}%
  \BibitemOpen
  \bibfield  {author} {\bibinfo {author} {\bibfnamefont {J.}~\bibnamefont
  {Liu}}, \bibinfo {author} {\bibfnamefont {X.}~\bibnamefont {Chen}}, \ and\
  \bibinfo {author} {\bibfnamefont {X.}~\bibnamefont {Ji}},\ }\href {\doibase
  10.1038/nphys4039} {\bibfield  {journal} {\bibinfo  {journal} {Nature Phys.}\
  }\textbf {\bibinfo {volume} {13}},\ \bibinfo {pages} {212} (\bibinfo {year}
  {2017})},\ \Eprint {http://arxiv.org/abs/1709.00688} {arXiv:1709.00688
  [astro-ph.CO]} \BibitemShut {NoStop}%
\bibitem [{\citenamefont {Moore}\ \emph
  {et~al.}(1999{\natexlab{a}})\citenamefont {Moore}, \citenamefont {Quinn},
  \citenamefont {Governato}, \citenamefont {Stadel},\ and\ \citenamefont
  {Lake}}]{Moore:1999gc}%
  \BibitemOpen
  \bibfield  {author} {\bibinfo {author} {\bibfnamefont {B.}~\bibnamefont
  {Moore}}, \bibinfo {author} {\bibfnamefont {T.~R.}\ \bibnamefont {Quinn}},
  \bibinfo {author} {\bibfnamefont {F.}~\bibnamefont {Governato}}, \bibinfo
  {author} {\bibfnamefont {J.}~\bibnamefont {Stadel}}, \ and\ \bibinfo {author}
  {\bibfnamefont {G.}~\bibnamefont {Lake}},\ }\href {\doibase
  10.1046/j.1365-8711.1999.03039.x} {\bibfield  {journal} {\bibinfo  {journal}
  {Mon. Not. Roy. Astron. Soc.}\ }\textbf {\bibinfo {volume} {310}},\ \bibinfo
  {pages} {1147} (\bibinfo {year} {1999}{\natexlab{a}})},\ \Eprint
  {http://arxiv.org/abs/astro-ph/9903164} {arXiv:astro-ph/9903164} \BibitemShut
  {NoStop}%
\bibitem [{\citenamefont {Klypin}\ \emph {et~al.}(1999)\citenamefont {Klypin},
  \citenamefont {Kravtsov}, \citenamefont {Valenzuela},\ and\ \citenamefont
  {Prada}}]{Klypin:1999uc}%
  \BibitemOpen
  \bibfield  {author} {\bibinfo {author} {\bibfnamefont {A.~A.}\ \bibnamefont
  {Klypin}}, \bibinfo {author} {\bibfnamefont {A.~V.}\ \bibnamefont
  {Kravtsov}}, \bibinfo {author} {\bibfnamefont {O.}~\bibnamefont
  {Valenzuela}}, \ and\ \bibinfo {author} {\bibfnamefont {F.}~\bibnamefont
  {Prada}},\ }\href {\doibase 10.1086/307643} {\bibfield  {journal} {\bibinfo
  {journal} {Astrophys. J.}\ }\textbf {\bibinfo {volume} {522}},\ \bibinfo
  {pages} {82} (\bibinfo {year} {1999})},\ \Eprint
  {http://arxiv.org/abs/astro-ph/9901240} {arXiv:astro-ph/9901240} \BibitemShut
  {NoStop}%
\bibitem [{\citenamefont {Moore}\ \emph
  {et~al.}(1999{\natexlab{b}})\citenamefont {Moore}, \citenamefont {Ghigna},
  \citenamefont {Governato}, \citenamefont {Lake}, \citenamefont {Quinn},
  \citenamefont {Stadel},\ and\ \citenamefont {Tozzi}}]{Moore:1999nt}%
  \BibitemOpen
  \bibfield  {author} {\bibinfo {author} {\bibfnamefont {B.}~\bibnamefont
  {Moore}}, \bibinfo {author} {\bibfnamefont {S.}~\bibnamefont {Ghigna}},
  \bibinfo {author} {\bibfnamefont {F.}~\bibnamefont {Governato}}, \bibinfo
  {author} {\bibfnamefont {G.}~\bibnamefont {Lake}}, \bibinfo {author}
  {\bibfnamefont {T.~R.}\ \bibnamefont {Quinn}}, \bibinfo {author}
  {\bibfnamefont {J.}~\bibnamefont {Stadel}}, \ and\ \bibinfo {author}
  {\bibfnamefont {P.}~\bibnamefont {Tozzi}},\ }\href {\doibase 10.1086/312287}
  {\bibfield  {journal} {\bibinfo  {journal} {Astrophys. J. Lett.}\ }\textbf
  {\bibinfo {volume} {524}},\ \bibinfo {pages} {L19} (\bibinfo {year}
  {1999}{\natexlab{b}})},\ \Eprint {http://arxiv.org/abs/astro-ph/9907411}
  {arXiv:astro-ph/9907411} \BibitemShut {NoStop}%
\bibitem [{\citenamefont {Boylan-Kolchin}\ \emph {et~al.}(2011)\citenamefont
  {Boylan-Kolchin}, \citenamefont {Bullock},\ and\ \citenamefont
  {Kaplinghat}}]{Boylan-Kolchin:2011qkt}%
  \BibitemOpen
  \bibfield  {author} {\bibinfo {author} {\bibfnamefont {M.}~\bibnamefont
  {Boylan-Kolchin}}, \bibinfo {author} {\bibfnamefont {J.~S.}\ \bibnamefont
  {Bullock}}, \ and\ \bibinfo {author} {\bibfnamefont {M.}~\bibnamefont
  {Kaplinghat}},\ }\href {\doibase 10.1111/j.1745-3933.2011.01074.x} {\bibfield
   {journal} {\bibinfo  {journal} {Mon. Not. Roy. Astron. Soc.}\ }\textbf
  {\bibinfo {volume} {415}},\ \bibinfo {pages} {L40} (\bibinfo {year}
  {2011})},\ \Eprint {http://arxiv.org/abs/1103.0007} {arXiv:1103.0007
  [astro-ph.CO]} \BibitemShut {NoStop}%
\bibitem [{\citenamefont {Weinberg}\ \emph {et~al.}(2015)\citenamefont
  {Weinberg}, \citenamefont {Bullock}, \citenamefont {Governato}, \citenamefont
  {Kuzio~de Naray},\ and\ \citenamefont {Peter}}]{Weinberg:2013aya}%
  \BibitemOpen
  \bibfield  {author} {\bibinfo {author} {\bibfnamefont {D.~H.}\ \bibnamefont
  {Weinberg}}, \bibinfo {author} {\bibfnamefont {J.~S.}\ \bibnamefont
  {Bullock}}, \bibinfo {author} {\bibfnamefont {F.}~\bibnamefont {Governato}},
  \bibinfo {author} {\bibfnamefont {R.}~\bibnamefont {Kuzio~de Naray}}, \ and\
  \bibinfo {author} {\bibfnamefont {A.~H.~G.}\ \bibnamefont {Peter}},\ }\href
  {\doibase 10.1073/pnas.1308716112} {\bibfield  {journal} {\bibinfo  {journal}
  {Proc. Nat. Acad. Sci.}\ }\textbf {\bibinfo {volume} {112}},\ \bibinfo
  {pages} {12249} (\bibinfo {year} {2015})},\ \Eprint
  {http://arxiv.org/abs/1306.0913} {arXiv:1306.0913 [astro-ph.CO]} \BibitemShut
  {NoStop}%
\bibitem [{\citenamefont {Hall}\ \emph {et~al.}(2010)\citenamefont {Hall},
  \citenamefont {Jedamzik}, \citenamefont {March-Russell},\ and\ \citenamefont
  {West}}]{Hall:2009bx}%
  \BibitemOpen
  \bibfield  {author} {\bibinfo {author} {\bibfnamefont {L.~J.}\ \bibnamefont
  {Hall}}, \bibinfo {author} {\bibfnamefont {K.}~\bibnamefont {Jedamzik}},
  \bibinfo {author} {\bibfnamefont {J.}~\bibnamefont {March-Russell}}, \ and\
  \bibinfo {author} {\bibfnamefont {S.~M.}\ \bibnamefont {West}},\ }\href
  {\doibase 10.1007/JHEP03(2010)080} {\bibfield  {journal} {\bibinfo  {journal}
  {JHEP}\ }\textbf {\bibinfo {volume} {03}},\ \bibinfo {pages} {080} (\bibinfo
  {year} {2010})},\ \Eprint {http://arxiv.org/abs/0911.1120} {arXiv:0911.1120
  [hep-ph]} \BibitemShut {NoStop}%
\bibitem [{\citenamefont {Essig}\ \emph {et~al.}(2012)\citenamefont {Essig},
  \citenamefont {Mardon},\ and\ \citenamefont {Volansky}}]{Essig:2011nj}%
  \BibitemOpen
  \bibfield  {author} {\bibinfo {author} {\bibfnamefont {R.}~\bibnamefont
  {Essig}}, \bibinfo {author} {\bibfnamefont {J.}~\bibnamefont {Mardon}}, \
  and\ \bibinfo {author} {\bibfnamefont {T.}~\bibnamefont {Volansky}},\ }\href
  {\doibase 10.1103/PhysRevD.85.076007} {\bibfield  {journal} {\bibinfo
  {journal} {Phys. Rev. D}\ }\textbf {\bibinfo {volume} {85}},\ \bibinfo
  {pages} {076007} (\bibinfo {year} {2012})},\ \Eprint
  {http://arxiv.org/abs/1108.5383} {arXiv:1108.5383 [hep-ph]} \BibitemShut
  {NoStop}%
\bibitem [{\citenamefont {Knapen}\ \emph {et~al.}(2017)\citenamefont {Knapen},
  \citenamefont {Lin},\ and\ \citenamefont {Zurek}}]{Knapen:2017xzo}%
  \BibitemOpen
  \bibfield  {author} {\bibinfo {author} {\bibfnamefont {S.}~\bibnamefont
  {Knapen}}, \bibinfo {author} {\bibfnamefont {T.}~\bibnamefont {Lin}}, \ and\
  \bibinfo {author} {\bibfnamefont {K.~M.}\ \bibnamefont {Zurek}},\ }\href
  {\doibase 10.1103/PhysRevD.96.115021} {\bibfield  {journal} {\bibinfo
  {journal} {Phys. Rev. D}\ }\textbf {\bibinfo {volume} {96}},\ \bibinfo
  {pages} {115021} (\bibinfo {year} {2017})},\ \Eprint
  {http://arxiv.org/abs/1709.07882} {arXiv:1709.07882 [hep-ph]} \BibitemShut
  {NoStop}%
\bibitem [{\citenamefont {Marsh}(2016)}]{Marsh:2015xka}%
  \BibitemOpen
  \bibfield  {author} {\bibinfo {author} {\bibfnamefont {D.~J.~E.}\
  \bibnamefont {Marsh}},\ }\href {\doibase 10.1016/j.physrep.2016.06.005}
  {\bibfield  {journal} {\bibinfo  {journal} {Phys. Rept.}\ }\textbf {\bibinfo
  {volume} {643}},\ \bibinfo {pages} {1} (\bibinfo {year} {2016})},\ \Eprint
  {http://arxiv.org/abs/1510.07633} {arXiv:1510.07633 [astro-ph.CO]}
  \BibitemShut {NoStop}%
\bibitem [{\citenamefont {An}\ \emph {et~al.}(2015)\citenamefont {An},
  \citenamefont {Pospelov}, \citenamefont {Pradler},\ and\ \citenamefont
  {Ritz}}]{An:2014twa}%
  \BibitemOpen
  \bibfield  {author} {\bibinfo {author} {\bibfnamefont {H.}~\bibnamefont
  {An}}, \bibinfo {author} {\bibfnamefont {M.}~\bibnamefont {Pospelov}},
  \bibinfo {author} {\bibfnamefont {J.}~\bibnamefont {Pradler}}, \ and\
  \bibinfo {author} {\bibfnamefont {A.}~\bibnamefont {Ritz}},\ }\href {\doibase
  10.1016/j.physletb.2015.06.018} {\bibfield  {journal} {\bibinfo  {journal}
  {Phys. Lett. B}\ }\textbf {\bibinfo {volume} {747}},\ \bibinfo {pages} {331}
  (\bibinfo {year} {2015})},\ \Eprint {http://arxiv.org/abs/1412.8378}
  {arXiv:1412.8378 [hep-ph]} \BibitemShut {NoStop}%
\bibitem [{\citenamefont {Pospelov}\ \emph {et~al.}(2008)\citenamefont
  {Pospelov}, \citenamefont {Ritz},\ and\ \citenamefont
  {Voloshin}}]{Pospelov:2008jk}%
  \BibitemOpen
  \bibfield  {author} {\bibinfo {author} {\bibfnamefont {M.}~\bibnamefont
  {Pospelov}}, \bibinfo {author} {\bibfnamefont {A.}~\bibnamefont {Ritz}}, \
  and\ \bibinfo {author} {\bibfnamefont {M.~B.}\ \bibnamefont {Voloshin}},\
  }\href {\doibase 10.1103/PhysRevD.78.115012} {\bibfield  {journal} {\bibinfo
  {journal} {Phys. Rev. D}\ }\textbf {\bibinfo {volume} {78}},\ \bibinfo
  {pages} {115012} (\bibinfo {year} {2008})},\ \Eprint
  {http://arxiv.org/abs/0807.3279} {arXiv:0807.3279 [hep-ph]} \BibitemShut
  {NoStop}%
\bibitem [{\citenamefont {Aprile}\ \emph {et~al.}(2017)\citenamefont {Aprile}
  \emph {et~al.}}]{XENON100:2017pfn}%
  \BibitemOpen
  \bibfield  {author} {\bibinfo {author} {\bibfnamefont {E.}~\bibnamefont
  {Aprile}} \emph {et~al.} (\bibinfo {collaboration} {XENON100}),\ }\href
  {\doibase 10.1103/PhysRevD.96.122002} {\bibfield  {journal} {\bibinfo
  {journal} {Phys. Rev. D}\ }\textbf {\bibinfo {volume} {96}},\ \bibinfo
  {pages} {122002} (\bibinfo {year} {2017})},\ \Eprint
  {http://arxiv.org/abs/1709.02222} {arXiv:1709.02222 [astro-ph.CO]}
  \BibitemShut {NoStop}%
\bibitem [{\citenamefont {Aprile}\ \emph {et~al.}(2022)\citenamefont {Aprile}
  \emph {et~al.}}]{XENON:2022ltv}%
  \BibitemOpen
  \bibfield  {author} {\bibinfo {author} {\bibfnamefont {E.}~\bibnamefont
  {Aprile}} \emph {et~al.} (\bibinfo {collaboration} {XENON}),\ }\href
  {\doibase 10.1103/PhysRevLett.129.161805} {\bibfield  {journal} {\bibinfo
  {journal} {Phys. Rev. Lett.}\ }\textbf {\bibinfo {volume} {129}},\ \bibinfo
  {pages} {161805} (\bibinfo {year} {2022})},\ \Eprint
  {http://arxiv.org/abs/2207.11330} {arXiv:2207.11330 [hep-ex]} \BibitemShut
  {NoStop}%
\bibitem [{\citenamefont {Agostini}\ \emph {et~al.}(2024)\citenamefont
  {Agostini} \emph {et~al.}}]{GERDA:2024gip}%
  \BibitemOpen
  \bibfield  {author} {\bibinfo {author} {\bibfnamefont {M.}~\bibnamefont
  {Agostini}} \emph {et~al.} (\bibinfo {collaboration} {GERDA}),\ }\href
  {\doibase 10.1140/epjc/s10052-024-13020-0} {\bibfield  {journal} {\bibinfo
  {journal} {Eur. Phys. J. C}\ }\textbf {\bibinfo {volume} {84}},\ \bibinfo
  {pages} {940} (\bibinfo {year} {2024})},\ \Eprint
  {http://arxiv.org/abs/2405.15954} {arXiv:2405.15954 [nucl-ex]} \BibitemShut
  {NoStop}%
\bibitem [{\citenamefont {Adhikari}\ \emph {et~al.}(2023)\citenamefont
  {Adhikari} \emph {et~al.}}]{COSINE-100:2023dir}%
  \BibitemOpen
  \bibfield  {author} {\bibinfo {author} {\bibfnamefont {G.}~\bibnamefont
  {Adhikari}} \emph {et~al.} (\bibinfo {collaboration} {COSINE-100}),\ }\href
  {\doibase 10.1103/PhysRevD.108.L041301} {\bibfield  {journal} {\bibinfo
  {journal} {Phys. Rev. D}\ }\textbf {\bibinfo {volume} {108}},\ \bibinfo
  {pages} {L041301} (\bibinfo {year} {2023})},\ \Eprint
  {http://arxiv.org/abs/2304.01460} {arXiv:2304.01460 [hep-ex]} \BibitemShut
  {NoStop}%
\bibitem [{\citenamefont {Abe}\ \emph {et~al.}(2018)\citenamefont {Abe} \emph
  {et~al.}}]{XMASS:2018pvs}%
  \BibitemOpen
  \bibfield  {author} {\bibinfo {author} {\bibfnamefont {K.}~\bibnamefont
  {Abe}} \emph {et~al.} (\bibinfo {collaboration} {XMASS}),\ }\href {\doibase
  10.1016/j.physletb.2018.10.050} {\bibfield  {journal} {\bibinfo  {journal}
  {Phys. Lett. B}\ }\textbf {\bibinfo {volume} {787}},\ \bibinfo {pages} {153}
  (\bibinfo {year} {2018})},\ \Eprint {http://arxiv.org/abs/1807.08516}
  {arXiv:1807.08516 [astro-ph.CO]} \BibitemShut {NoStop}%
\bibitem [{\citenamefont {Akerib}\ \emph {et~al.}(2017)\citenamefont {Akerib}
  \emph {et~al.}}]{LUX:2017glr}%
  \BibitemOpen
  \bibfield  {author} {\bibinfo {author} {\bibfnamefont {D.~S.}\ \bibnamefont
  {Akerib}} \emph {et~al.} (\bibinfo {collaboration} {LUX}),\ }\href {\doibase
  10.1103/PhysRevLett.118.261301} {\bibfield  {journal} {\bibinfo  {journal}
  {Phys. Rev. Lett.}\ }\textbf {\bibinfo {volume} {118}},\ \bibinfo {pages}
  {261301} (\bibinfo {year} {2017})},\ \Eprint
  {http://arxiv.org/abs/1704.02297} {arXiv:1704.02297 [astro-ph.CO]}
  \BibitemShut {NoStop}%
\bibitem [{\citenamefont {Fu}\ \emph {et~al.}(2017)\citenamefont {Fu} \emph
  {et~al.}}]{PandaX:2017ock}%
  \BibitemOpen
  \bibfield  {author} {\bibinfo {author} {\bibfnamefont {C.}~\bibnamefont {Fu}}
  \emph {et~al.} (\bibinfo {collaboration} {PandaX}),\ }\href {\doibase
  10.1103/PhysRevLett.119.181806} {\bibfield  {journal} {\bibinfo  {journal}
  {Phys. Rev. Lett.}\ }\textbf {\bibinfo {volume} {119}},\ \bibinfo {pages}
  {181806} (\bibinfo {year} {2017})},\ \Eprint
  {http://arxiv.org/abs/1707.07921} {arXiv:1707.07921 [hep-ex]} \BibitemShut
  {NoStop}%
\bibitem [{\citenamefont {Armengaud}\ \emph {et~al.}(2018)\citenamefont
  {Armengaud} \emph {et~al.}}]{EDELWEISS:2018tde}%
  \BibitemOpen
  \bibfield  {author} {\bibinfo {author} {\bibfnamefont {E.}~\bibnamefont
  {Armengaud}} \emph {et~al.} (\bibinfo {collaboration} {EDELWEISS}),\ }\href
  {\doibase 10.1103/PhysRevD.98.082004} {\bibfield  {journal} {\bibinfo
  {journal} {Phys. Rev. D}\ }\textbf {\bibinfo {volume} {98}},\ \bibinfo
  {pages} {082004} (\bibinfo {year} {2018})},\ \Eprint
  {http://arxiv.org/abs/1808.02340} {arXiv:1808.02340 [hep-ex]} \BibitemShut
  {NoStop}%
\bibitem [{\citenamefont {Arnquist}\ \emph {et~al.}(2024)\citenamefont
  {Arnquist} \emph {et~al.}}]{Majorana:2022gtu}%
  \BibitemOpen
  \bibfield  {author} {\bibinfo {author} {\bibfnamefont {I.~J.}\ \bibnamefont
  {Arnquist}} \emph {et~al.} (\bibinfo {collaboration} {Majorana}),\ }\href
  {\doibase 10.1103/PhysRevLett.132.041001} {\bibfield  {journal} {\bibinfo
  {journal} {Phys. Rev. Lett.}\ }\textbf {\bibinfo {volume} {132}},\ \bibinfo
  {pages} {041001} (\bibinfo {year} {2024})},\ \Eprint
  {http://arxiv.org/abs/2206.10638} {arXiv:2206.10638 [hep-ex]} \BibitemShut
  {NoStop}%
\bibitem [{\citenamefont {Aalbers}\ \emph {et~al.}(2023)\citenamefont {Aalbers}
  \emph {et~al.}}]{LZ:2023poo}%
  \BibitemOpen
  \bibfield  {author} {\bibinfo {author} {\bibfnamefont {J.}~\bibnamefont
  {Aalbers}} \emph {et~al.} (\bibinfo {collaboration} {LZ}),\ }\href {\doibase
  10.1103/PhysRevD.108.072006} {\bibfield  {journal} {\bibinfo  {journal}
  {Phys. Rev. D}\ }\textbf {\bibinfo {volume} {108}},\ \bibinfo {pages}
  {072006} (\bibinfo {year} {2023})},\ \Eprint
  {http://arxiv.org/abs/2307.15753} {arXiv:2307.15753 [hep-ex]} \BibitemShut
  {NoStop}%
\bibitem [{\citenamefont {Ferreira}\ \emph {et~al.}(2022)\citenamefont
  {Ferreira}, \citenamefont {Marsh},\ and\ \citenamefont
  {M\"uller}}]{Ferreira:2022egk}%
  \BibitemOpen
  \bibfield  {author} {\bibinfo {author} {\bibfnamefont {R.~Z.}\ \bibnamefont
  {Ferreira}}, \bibinfo {author} {\bibfnamefont {M.~C.~D.}\ \bibnamefont
  {Marsh}}, \ and\ \bibinfo {author} {\bibfnamefont {E.}~\bibnamefont
  {M\"uller}},\ }\href {\doibase 10.1103/PhysRevLett.128.221302} {\bibfield
  {journal} {\bibinfo  {journal} {Phys. Rev. Lett.}\ }\textbf {\bibinfo
  {volume} {128}},\ \bibinfo {pages} {221302} (\bibinfo {year} {2022})},\
  \Eprint {http://arxiv.org/abs/2202.08858} {arXiv:2202.08858 [hep-ph]}
  \BibitemShut {NoStop}%
\bibitem [{\citenamefont {Si}\ \emph {et~al.}(2022)\citenamefont {Si} \emph
  {et~al.}}]{PandaX:2022kwg}%
  \BibitemOpen
  \bibfield  {author} {\bibinfo {author} {\bibfnamefont {L.}~\bibnamefont {Si}}
  \emph {et~al.} (\bibinfo {collaboration} {PandaX}),\ }\href {\doibase
  10.34133/2022/9798721} {\bibfield  {journal} {\bibinfo  {journal} {Research}\
  }\textbf {\bibinfo {volume} {2022}},\ \bibinfo {pages} {9798721} (\bibinfo
  {year} {2022})},\ \Eprint {http://arxiv.org/abs/2205.12809} {arXiv:2205.12809
  [nucl-ex]} \BibitemShut {NoStop}%
\bibitem [{\citenamefont {Yan}\ \emph {et~al.}(2024)\citenamefont {Yan} \emph
  {et~al.}}]{PandaX:2023ggs}%
  \BibitemOpen
  \bibfield  {author} {\bibinfo {author} {\bibfnamefont {X.}~\bibnamefont
  {Yan}} \emph {et~al.} (\bibinfo {collaboration} {PandaX}),\ }\href {\doibase
  10.1103/PhysRevLett.132.152502} {\bibfield  {journal} {\bibinfo  {journal}
  {Phys. Rev. Lett.}\ }\textbf {\bibinfo {volume} {132}},\ \bibinfo {pages}
  {152502} (\bibinfo {year} {2024})},\ \Eprint
  {http://arxiv.org/abs/2312.15632} {arXiv:2312.15632 [nucl-ex]} \BibitemShut
  {NoStop}%
\bibitem [{\citenamefont {Meng}\ \emph {et~al.}(2021)\citenamefont {Meng} \emph
  {et~al.}}]{PandaX-4T:2021bab}%
  \BibitemOpen
  \bibfield  {author} {\bibinfo {author} {\bibfnamefont {Y.}~\bibnamefont
  {Meng}} \emph {et~al.} (\bibinfo {collaboration} {PandaX-4T}),\ }\href
  {\doibase 10.1103/PhysRevLett.127.261802} {\bibfield  {journal} {\bibinfo
  {journal} {Phys. Rev. Lett.}\ }\textbf {\bibinfo {volume} {127}},\ \bibinfo
  {pages} {261802} (\bibinfo {year} {2021})},\ \Eprint
  {http://arxiv.org/abs/2107.13438} {arXiv:2107.13438 [hep-ex]} \BibitemShut
  {NoStop}%
\bibitem [{\citenamefont {Lu}\ \emph {et~al.}(2024)\citenamefont {Lu} \emph
  {et~al.}}]{Lu:2024ilt}%
  \BibitemOpen
  \bibfield  {author} {\bibinfo {author} {\bibfnamefont {X.}~\bibnamefont {Lu}}
  \emph {et~al.},\ }\href@noop {} {\bibfield  {journal} {\bibinfo  {journal}
  {Chinese Physics C}\ } (\bibinfo {year} {2024})},\ \Eprint
  {http://arxiv.org/abs/2401.07045} {arXiv:2401.07045 [hep-ex]} \BibitemShut
  {NoStop}%
\bibitem [{\citenamefont {Luo}\ \emph {et~al.}(2024)\citenamefont {Luo} \emph
  {et~al.}}]{PandaX:2024med}%
  \BibitemOpen
  \bibfield  {author} {\bibinfo {author} {\bibfnamefont {Y.}~\bibnamefont
  {Luo}} \emph {et~al.} (\bibinfo {collaboration} {PandaX}),\ }\href {\doibase
  10.1103/PhysRevD.110.023029} {\bibfield  {journal} {\bibinfo  {journal}
  {Phys. Rev. D}\ }\textbf {\bibinfo {volume} {110}},\ \bibinfo {pages}
  {023029} (\bibinfo {year} {2024})},\ \Eprint
  {http://arxiv.org/abs/2403.04239} {arXiv:2403.04239 [physics.ins-det]}
  \BibitemShut {NoStop}%
\bibitem [{\citenamefont {Szydagis}\ \emph
  {et~al.}(2011{\natexlab{a}})\citenamefont {Szydagis}, \citenamefont {Barry},
  \citenamefont {Kazkaz}, \citenamefont {Mock}, \citenamefont {Stolp},
  \citenamefont {Sweany}, \citenamefont {Tripathi}, \citenamefont {Uvarov},
  \citenamefont {Walsh},\ and\ \citenamefont {Woods}}]{Szydagis:2011tk}%
  \BibitemOpen
  \bibfield  {author} {\bibinfo {author} {\bibfnamefont {M.}~\bibnamefont
  {Szydagis}}, \bibinfo {author} {\bibfnamefont {N.}~\bibnamefont {Barry}},
  \bibinfo {author} {\bibfnamefont {K.}~\bibnamefont {Kazkaz}}, \bibinfo
  {author} {\bibfnamefont {J.}~\bibnamefont {Mock}}, \bibinfo {author}
  {\bibfnamefont {D.}~\bibnamefont {Stolp}}, \bibinfo {author} {\bibfnamefont
  {M.}~\bibnamefont {Sweany}}, \bibinfo {author} {\bibfnamefont
  {M.}~\bibnamefont {Tripathi}}, \bibinfo {author} {\bibfnamefont
  {S.}~\bibnamefont {Uvarov}}, \bibinfo {author} {\bibfnamefont
  {N.}~\bibnamefont {Walsh}}, \ and\ \bibinfo {author} {\bibfnamefont
  {M.}~\bibnamefont {Woods}},\ }\href {\doibase 10.1088/1748-0221/6/10/P10002}
  {\bibfield  {journal} {\bibinfo  {journal} {JINST}\ }\textbf {\bibinfo
  {volume} {6}},\ \bibinfo {pages} {P10002} (\bibinfo {year}
  {2011}{\natexlab{a}})},\ \Eprint {http://arxiv.org/abs/1106.1613}
  {arXiv:1106.1613 [physics.ins-det]} \BibitemShut {NoStop}%
\bibitem [{\citenamefont {Bo}\ \emph {et~al.}(2024{\natexlab{a}})\citenamefont
  {Bo} \emph {et~al.}}]{PandaX:2024xpq}%
  \BibitemOpen
  \bibfield  {author} {\bibinfo {author} {\bibfnamefont {Z.}~\bibnamefont {Bo}}
  \emph {et~al.} (\bibinfo {collaboration} {PandaX}),\ }\href@noop {} {\
  (\bibinfo {year} {2024}{\natexlab{a}})},\ \Eprint
  {http://arxiv.org/abs/2411.14355} {arXiv:2411.14355 [nucl-ex]} \BibitemShut
  {NoStop}%
\bibitem [{\citenamefont {Chen}\ \emph {et~al.}(2021)\citenamefont {Chen} \emph
  {et~al.}}]{Chen:2021asx}%
  \BibitemOpen
  \bibfield  {author} {\bibinfo {author} {\bibfnamefont {X.}~\bibnamefont
  {Chen}} \emph {et~al.},\ }\href {\doibase 10.1088/1748-0221/16/09/T09004}
  {\bibfield  {journal} {\bibinfo  {journal} {JINST}\ }\textbf {\bibinfo
  {volume} {16}},\ \bibinfo {pages} {T09004} (\bibinfo {year} {2021})},\
  \Eprint {http://arxiv.org/abs/2107.05935} {arXiv:2107.05935
  [physics.ins-det]} \BibitemShut {NoStop}%
\bibitem [{\citenamefont {Szydagis}\ \emph
  {et~al.}(2011{\natexlab{b}})\citenamefont {Szydagis}, \citenamefont {Barry},
  \citenamefont {Kazkaz}, \citenamefont {Mock}, \citenamefont {Stolp},
  \citenamefont {Sweany}, \citenamefont {Tripathi}, \citenamefont {Uvarov},
  \citenamefont {Walsh},\ and\ \citenamefont {Woods}}]{szydagis2011nest}%
  \BibitemOpen
  \bibfield  {author} {\bibinfo {author} {\bibfnamefont {M.}~\bibnamefont
  {Szydagis}}, \bibinfo {author} {\bibfnamefont {N.}~\bibnamefont {Barry}},
  \bibinfo {author} {\bibfnamefont {K.}~\bibnamefont {Kazkaz}}, \bibinfo
  {author} {\bibfnamefont {J.}~\bibnamefont {Mock}}, \bibinfo {author}
  {\bibfnamefont {D.}~\bibnamefont {Stolp}}, \bibinfo {author} {\bibfnamefont
  {M.}~\bibnamefont {Sweany}}, \bibinfo {author} {\bibfnamefont
  {M.}~\bibnamefont {Tripathi}}, \bibinfo {author} {\bibfnamefont
  {S.}~\bibnamefont {Uvarov}}, \bibinfo {author} {\bibfnamefont
  {N.}~\bibnamefont {Walsh}}, \ and\ \bibinfo {author} {\bibfnamefont
  {M.}~\bibnamefont {Woods}},\ }\href@noop {} {\bibfield  {journal} {\bibinfo
  {journal} {Journal of Instrumentation}\ }\textbf {\bibinfo {volume} {6}},\
  \bibinfo {pages} {P10002} (\bibinfo {year} {2011}{\natexlab{b}})}\BibitemShut
  {NoStop}%
\bibitem [{\citenamefont {Bo}\ \emph {et~al.}(2024{\natexlab{b}})\citenamefont
  {Bo} \emph {et~al.}}]{PandaX:2024qfu}%
  \BibitemOpen
  \bibfield  {author} {\bibinfo {author} {\bibfnamefont {Z.}~\bibnamefont {Bo}}
  \emph {et~al.} (\bibinfo {collaboration} {PandaX}),\ }\href@noop {} {\
  (\bibinfo {year} {2024}{\natexlab{b}})},\ \Eprint
  {http://arxiv.org/abs/2408.00664} {arXiv:2408.00664 [hep-ex]} \BibitemShut
  {NoStop}%
\bibitem [{\citenamefont {Collon}\ \emph {et~al.}(2004)\citenamefont {Collon},
  \citenamefont {Kutschera},\ and\ \citenamefont {Lu}}]{Collon:2004xs}%
  \BibitemOpen
  \bibfield  {author} {\bibinfo {author} {\bibfnamefont {P.}~\bibnamefont
  {Collon}}, \bibinfo {author} {\bibfnamefont {W.}~\bibnamefont {Kutschera}}, \
  and\ \bibinfo {author} {\bibfnamefont {Z.-T.}\ \bibnamefont {Lu}},\ }\href
  {\doibase 10.1146/annurev.nucl.53.041002.110622} {\bibfield  {journal}
  {\bibinfo  {journal} {Ann. Rev. Nucl. Part. Sci.}\ }\textbf {\bibinfo
  {volume} {54}},\ \bibinfo {pages} {39} (\bibinfo {year} {2004})},\ \Eprint
  {http://arxiv.org/abs/nucl-ex/0402013} {arXiv:nucl-ex/0402013} \BibitemShut
  {NoStop}%
\bibitem [{\citenamefont {Chen}\ \emph {et~al.}(2017)\citenamefont {Chen},
  \citenamefont {Chi}, \citenamefont {Liu},\ and\ \citenamefont
  {Wu}}]{Chen:2016eab}%
  \BibitemOpen
  \bibfield  {author} {\bibinfo {author} {\bibfnamefont {J.-W.}\ \bibnamefont
  {Chen}}, \bibinfo {author} {\bibfnamefont {H.-C.}\ \bibnamefont {Chi}},
  \bibinfo {author} {\bibfnamefont {C.~P.}\ \bibnamefont {Liu}}, \ and\
  \bibinfo {author} {\bibfnamefont {C.-P.}\ \bibnamefont {Wu}},\ }\href
  {\doibase 10.1016/j.physletb.2017.10.029} {\bibfield  {journal} {\bibinfo
  {journal} {Phys. Lett. B}\ }\textbf {\bibinfo {volume} {774}},\ \bibinfo
  {pages} {656} (\bibinfo {year} {2017})},\ \Eprint
  {http://arxiv.org/abs/1610.04177} {arXiv:1610.04177 [hep-ex]} \BibitemShut
  {NoStop}%
\bibitem [{\citenamefont {Bellini}\ \emph {et~al.}(2014)\citenamefont {Bellini}
  \emph {et~al.}}]{BOREXINO:2014pcl}%
  \BibitemOpen
  \bibfield  {author} {\bibinfo {author} {\bibfnamefont {G.}~\bibnamefont
  {Bellini}} \emph {et~al.} (\bibinfo {collaboration} {BOREXINO}),\ }\href
  {\doibase 10.1038/nature13702} {\bibfield  {journal} {\bibinfo  {journal}
  {Nature}\ }\textbf {\bibinfo {volume} {512}},\ \bibinfo {pages} {383}
  (\bibinfo {year} {2014})}\BibitemShut {NoStop}%
\bibitem [{\citenamefont {{National Nuclear Data Center}}()}]{nndc_nudat}%
  \BibitemOpen
  \bibfield  {author} {\bibinfo {author} {\bibnamefont {{National Nuclear Data
  Center}}},\ }\href@noop {} {\enquote {\bibinfo {title} {Information extracted
  from the nudat database},}\ }\bibinfo {howpublished}
  {\url{https://www.nndc.bnl.gov/nudat/}}\BibitemShut {NoStop}%
\bibitem [{\citenamefont {Ranucci}(2012)}]{Ranucci:2012ed}%
  \BibitemOpen
  \bibfield  {author} {\bibinfo {author} {\bibfnamefont {G.}~\bibnamefont
  {Ranucci}},\ }\href {\doibase 10.1016/j.nima.2011.09.047} {\bibfield
  {journal} {\bibinfo  {journal} {Nucl. Instrum. Meth. A}\ }\textbf {\bibinfo
  {volume} {661}},\ \bibinfo {pages} {77} (\bibinfo {year} {2012})},\ \Eprint
  {http://arxiv.org/abs/1201.4604} {arXiv:1201.4604 [physics.data-an]}
  \BibitemShut {NoStop}%
\bibitem [{\citenamefont {Cowan}\ \emph {et~al.}(2011)\citenamefont {Cowan},
  \citenamefont {Cranmer}, \citenamefont {Gross},\ and\ \citenamefont
  {Vitells}}]{Cowan:2010js}%
  \BibitemOpen
  \bibfield  {author} {\bibinfo {author} {\bibfnamefont {G.}~\bibnamefont
  {Cowan}}, \bibinfo {author} {\bibfnamefont {K.}~\bibnamefont {Cranmer}},
  \bibinfo {author} {\bibfnamefont {E.}~\bibnamefont {Gross}}, \ and\ \bibinfo
  {author} {\bibfnamefont {O.}~\bibnamefont {Vitells}},\ }\href {\doibase
  10.1140/epjc/s10052-011-1554-0} {\bibfield  {journal} {\bibinfo  {journal}
  {Eur. Phys. J. C}\ }\textbf {\bibinfo {volume} {71}},\ \bibinfo {pages}
  {1554} (\bibinfo {year} {2011})},\ \bibinfo {note} {[Erratum: Eur.Phys.J.C
  73, 2501 (2013)]},\ \Eprint {http://arxiv.org/abs/1007.1727} {arXiv:1007.1727
  [physics.data-an]} \BibitemShut {NoStop}%
\bibitem [{\citenamefont {Zeng}\ \emph {et~al.}(2024)\citenamefont {Zeng} \emph
  {et~al.}}]{PandaX:2024cic}%
  \BibitemOpen
  \bibfield  {author} {\bibinfo {author} {\bibfnamefont {X.}~\bibnamefont
  {Zeng}} \emph {et~al.} (\bibinfo {collaboration} {PandaX}),\ }\href@noop {}
  {\  (\bibinfo {year} {2024})},\ \Eprint {http://arxiv.org/abs/2408.07641}
  {arXiv:2408.07641 [hep-ex]} \BibitemShut {NoStop}%
\bibitem [{\citenamefont {Redondo}\ and\ \citenamefont
  {Postma}(2009)}]{Redondo:2008ec}%
  \BibitemOpen
  \bibfield  {author} {\bibinfo {author} {\bibfnamefont {J.}~\bibnamefont
  {Redondo}}\ and\ \bibinfo {author} {\bibfnamefont {M.}~\bibnamefont
  {Postma}},\ }\href {\doibase 10.1088/1475-7516/2009/02/005} {\bibfield
  {journal} {\bibinfo  {journal} {JCAP}\ }\textbf {\bibinfo {volume} {02}},\
  \bibinfo {pages} {005} (\bibinfo {year} {2009})},\ \Eprint
  {http://arxiv.org/abs/0811.0326} {arXiv:0811.0326 [hep-ph]} \BibitemShut
  {NoStop}%
\bibitem [{\citenamefont {Li}\ and\ \citenamefont {Xu}(2023)}]{Li:2023vpv}%
  \BibitemOpen
  \bibfield  {author} {\bibinfo {author} {\bibfnamefont {S.-P.}\ \bibnamefont
  {Li}}\ and\ \bibinfo {author} {\bibfnamefont {X.-J.}\ \bibnamefont {Xu}},\
  }\href {\doibase 10.1088/1475-7516/2023/09/009} {\bibfield  {journal}
  {\bibinfo  {journal} {JCAP}\ }\textbf {\bibinfo {volume} {09}},\ \bibinfo
  {pages} {009} (\bibinfo {year} {2023})},\ \Eprint
  {http://arxiv.org/abs/2304.12907} {arXiv:2304.12907 [hep-ph]} \BibitemShut
  {NoStop}%
\end{thebibliography}%
\end{document}